\documentclass[journal=langd5,manuscript=article]{achemso}

\usepackage[version=3]{mhchem} 
\usepackage{bm}
\usepackage{amsmath,amssymb}
\usepackage{txfonts}
\usepackage{type1cm}

\author{Masao Iwamatsu}
\email{iwamatsu@ph.ns.tcu.ac.jp}
\phone{+81 (0)3 5705 0104 ext. 2382}
\fax{+81 (0)3 5707 2222}
\affiliation[Tokyo City University]
{Department of Physics, Faculty of Liberal Arts and Sciences, Tokyo City University, Setagaya-ku, Tokyo 158-8557, JAPAN}

\title[Line tension in nucleation]{Line-tension effects on heterogeneous nucleation on a spherical substrate and in a spherical cavity}

\abbreviations{IR,NMR,UV}
\keywords{American Chemical Society, \LaTeX}

\begin{document}
\begin{abstract}
The line-tension effects on heterogeneous nucleation are considered when a spherical lens-shaped nucleus is nucleated on top of a spherical substrate and on the bottom of the wall of a spherical cavity.  The effect of line tension on the nucleation barrier can be separated from the usual volume term.  As the radius of the substrate increases, the nucleation barrier decreases and approaches that of a flat substrate.  However, as the radius of the cavity  increases, the nucleation barrier increases and approaches that of a flat substrate.  A small spherical substrate is a less active nucleation site than a flat substrate, and a small spherical cavity is a more active nucleation site than a flat substrate.  In contrast, the line-tension effect on the nucleation barrier is maximum when the radii of the nucleus and the substrate or cavity become comparable.  Therefore, by tuning the size of the spherical substrate or spherical cavity, the effect of the line tension can be optimized.  These results will be useful in broad range of applications from material processing to understanding of global climate, where the heterogeneous nucleation plays a vital role. 

\end{abstract}

\section{1. Introduction}
Nucleation is often heterogeneous and is assisted by the presence of a substrate or wall and impurities.  Although heterogeneous nucleation plays a vital role in the process of various material processing applications, ranging from steel production to food and beverage production~\cite{Kelton2010}, the theoretical study of this phenomenon has been hindered by the complexity of the spherical geometry.  In fact, the heterogeneous nucleation in a spherical cavity and on a spherical substrate is important in broad range of applications from bubble formation in carbonated drinks in daily life~\cite{Kelton2010} to the cloud formation which controls global climate~\cite{Hellmuth2013}.  Since a three-phase contact line is always involved in the heterogeneous nucleation, theoretical understanding of the line tension must be crucial~\cite{Hellmuth2013}.   However, most recent theoretical studies of heterogeneous nucleation rely on computer simulations~\cite{Sear2008,Liu2011,Sandomirski2014} or ad-hock assumptions~\cite{Ghosh2013} for a realistically constructed model.

The theoretical formulation of heterogeneous nucleation of a lens-shaped nucleus on an ideally spherical substrate was conducted half a century ago (in 1958) by Fletcher~\cite{Fletcher1958} following the work of Turnbull~\cite{Turnbull1950} on a flat substrate. The theory has been applied, for example, to study heterogeneous nucleation in the atmosphere~\cite{Lazaridis1991,Gorbunov1997}.  Further theoretical analysis has been necessary because of the mathematical complexity of the problem~\cite{Xu2005,Qian2007,Qian2009}. The same problem has also been studied from the standpoint of wetting~\cite{Hage1984}.  However, thus far, few studies of the line-tension effect on a spherical substrate have been conducted~\cite{Scheludko1985}, while many have been conducted on this phenomena on a flat substrate~\cite{Navascues1981,Widom1995,Greer2010}.  There are also several studies of heterogeneous nucleation in a confined volume~\cite{Cooper2007,Liu2008} and within a cavity~\cite{Maksimov2010,Maksimov2013}.  A similar problem of a macroscopic droplet on spherical convex and concave surfaces has also been studied~\cite{Extrand2012}.  However, these works do not focus much on the line tension~\cite{Greer2010}. 

In this paper, we will consider the heterogeneous nucleation of a lens-shaped critical nucleus on a spherical substrate and on a wall of a spherical cavity within the framework of classical nucleation theory (CNT).  We will regard the critical nucleus as a continuum with a uniform density and a sharp interface of a part of a spherical surface.  Therefore, the contact line and angle can be defined without ambiguity.  The nucleus-substrate interaction is confined to the contact area so that the classical concept of surface tension and interfacial energy can be applied.

\section{\label{sec:sec2}2. Line-tension effect on contact angle}

\subsection{A. Nucleus on a convex spherical substrate}

We will consider a lens-shaped liquid nucleus nucleated on a spherical substrate from oversaturated vapor.  However, the result is general and can be applied to the nucleation of crystal grains or vapor bubbles.  According to the classical idea of wetting and nucleation theory~\cite{Fletcher1958,Navascues1981,Qian2009}, the Helmholtz free energy of a nucleus (sessile droplet) is given by
\begin{equation}
\Delta F=\sigma_{\rm lv}A_{\rm lv}+\Delta\sigma A_{\rm sl}+\tau L,
\label{eq:N1}
\end{equation}
and
\begin{equation}
\Delta\sigma = \sigma_{\rm sl}-\sigma_{\rm sv},
\label{eq:N2}
\end{equation}
where $A_{\rm lv}$ and $A_{\rm sl}$ are the surface area of the liquid-vapor and liquid-solid (substrate) interfaces and their surface tensions are $\sigma_{\rm lv}$ and $\sigma_{\rm sl}$, respectively.  $\Delta \sigma$ is the free energy gain as the solid-vapor interface with the surface tension $\sigma_{\rm sv}$ is replaced by the solid-liquid interface with the surface tension $\sigma_{\rm sl}$.   The effect of the line tension $\tau$ is given by the last term, where $L$ denotes the length of the three-phase contact line.  Instead of using the standard grand potential which is relevant to the nucleation problem~\cite{Hellmuth2013} of volatile liquids, we use the Helmholtz free energy because we need to study the contact angle for fixed droplet volume.  In fact, the grand potential and the Helmholtz free energy are identical for fixed volume.  Therefore, the result is also useful to study the stability of droplet of nonvolatile liquid on a spherical substrate. 

\begin{figure}[htbp]
\begin{center}
\includegraphics[width=0.6\linewidth]{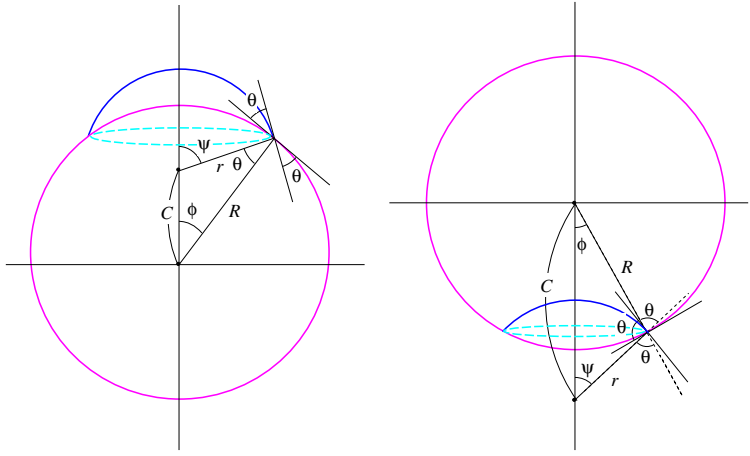}
\caption{
(a) A liquid nucleus (blue) on a convex spherical substrate (pink), and (b) that on an inner wall of a concave spherical cavity.  The three-phase contact line is indicated by light-blue broken lines.  The center of the nucleus with the radius $r$ and that of the spherical substrate or cavity with radius $R$ are separated by the distance $C$.  The contact angle is denoted by $\theta$. Note that the three-phase contact line passes through the equator when $\phi=90^{\circ}$ and the contact line moves from the upper (lower) hemisphere to lower (upper) hemisphere.  }
\label{fig:N1}
\end{center}
\end{figure}

We consider the semi-spherical nucleus with radius $r$ nucleated on a spherical substrate with radius $R$ shown in Figure~\ref{fig:N1}(a).  Using the geometrical angles $\phi$ and $\psi$ together with the distance $C$ between the centers of the two spheres, we obtain 
\begin{equation}
A_{\rm lv}=2\pi r^{2}\left(1-\cos\psi\right)=\pi r\frac{\left(C+r\right)^2-R^{2}}{C},
\label{eq:N3}
\end{equation}
and
\begin{equation}
A_{\rm sl}=2\pi R^{2}\left(1-\cos\phi\right)=\pi R\frac{r^{2}-\left(C-R\right)^2}{C}
\label{eq:N4}
\end{equation}
for the surface areas of the liquid-vapor and liquid-solid interfaces, respectively.  The length of the three-phase contact line is given by
\begin{equation}
L=2\pi R\sin\phi=2\pi r\sin\psi=\frac{2\pi Rr}{C}\sin\theta=\frac{\pi}{C}\sqrt{2C^{2}\left(R^{2}+r^{2}\right)-\left(R^{2}-r^{2}\right)^2-C^{4}},
\label{eq:N5}
\end{equation}
where $\theta$ is the contact angle, which is related to the angles $\phi$ and $\psi$ (Fig.~\ref{fig:N1}(a)) through 
\begin{eqnarray}
C\sin\phi &=& r\sin\theta,\;\;\;C\sin\psi=R\sin\theta, \nonumber \\
C\cos\phi &=& R-r\cos\theta,\;\;\;\; C\cos\psi=\pm\left(r-R\cos\theta\right),
\label{eq:N6}
\end{eqnarray}
and to the distance $C$ through
\begin{equation}
C^2=R^{2}+r^{2}-2Rr\cos\theta.
\label{eq:N7}
\end{equation}
Using Eqs.~(\ref{eq:N3}), (\ref{eq:N4}) and (\ref{eq:N5}), the Helmholtz free energy, Eq.~(\ref{eq:N1}), is given as a function of $C$ and $r$ for a given radius $R$ of the spherical substrate.

The volume of the nucleus can be obtained using the integration scheme originally developed by Hamaker~\cite{Hamaker1937},
\begin{eqnarray}
V &=& \int_{R}^{r+C}\frac{\pi R}{C}\left[ r^{2}-\left(C-R\right)^2\right]dR
\nonumber \\
&=&\frac{\pi}{12C}\left(C-R+r\right)^{2}\left[3\left(R+r\right)^2-2C\left(R-r\right)-C^{2}\right],
\label{eq:N8}
\end{eqnarray}
and is expressed by $R$, $r$ and $C$.  Eq.~(\ref{eq:N8}) reduces to the more intuitive formula
\begin{equation}
V=\frac{\pi r^3}{3}\left(2-3\cos\psi+\cos^{3}\psi\right)-\frac{\pi R^3}{3}\left(2-3\cos\phi+\cos^{3}\phi\right).
\label{eq:N9}
\end{equation}

The contact angle is determined by minimizing the Helmholtz free energy in Eq.~(\ref{eq:N1}) under the condition of a constant nucleus volume $V$.  Using Eq.~(\ref{eq:N8}), one can obtain
\begin{equation}
\frac{dC}{dr}=-\left(\frac{\partial V}{\partial r}\right)/\left(\frac{\partial V}{\partial C}\right)
\label{eq:N10}
\end{equation}
from $dV=0$, and the minimization of Eq.~(\ref{eq:N1}) under the condition of constant volume by
\begin{equation}
\frac{d\Delta F}{dr}=\frac{\partial \Delta F}{\partial r}+\frac{\partial \Delta F}{\partial C}\frac{dC}{dr}=0
\label{eq:N11}
\end{equation}
leads to the equation
\begin{equation}
\Delta \sigma=-\frac{R^{2}+r^{2}-C^{2}}{2Rr}\sigma_{\rm lv}-\frac{C^{2}+R^{2}-r^{2}}{R\sqrt{2C^{2}\left(R^{2}+r^{2}\right)-\left(R^{2}-r^{2}\right)^{2}-C^{4}}}\tau,
\label{eq:N12}
\end{equation}
which gives the contact angle $\theta$ through
\begin{equation}
\Delta\sigma+\sigma_{\rm lv}\cos\theta+\frac{R-r\cos\theta}{Rr\sin\theta}\tau=0,
\label{eq:N13}
\end{equation}
similar to the classical Young equation~\cite{Young1805} on a flat substrate,
\begin{equation}
\Delta\sigma+\sigma_{\rm lv}\cos\theta_{0}=0,
\label{eq:N14}
\end{equation}
where $\theta_{0}$ is the classical Young's contact angle.  Therefore, even on a spherical curved surface, the contact angle will be determined from the classical Young equation (\ref{eq:N14}) for flat surfaces~\cite{Fletcher1958,Qian2009} when the line tension can be neglected ($\tau=0$).

Equation (\ref{eq:N13}) can also be written as
\begin{equation}
\sigma_{\rm lv}\cos\theta=\sigma_{\rm lv}\cos\theta_{0}-\frac{\tau}{R\tan\phi},
\label{eq:N15}
\end{equation}
which is known as the generalized Young equation~\cite{Hienola2007}. This formula is different from that proposed by Scheludko~\cite{Scheludko1985}, but is the same as that proposed by Heinola~\cite{Hienola2007}, which was originally derived for the critical nucleus of heterogeneous nucleation; the radius $r$ is assumed to be given by the Young-Laplace formula.  In contrast, our derivation is based on the general principle of the minimization of the Helmholtz free energy, and the size $r$ takes any values determined from the volume of the nucleus.  Therefore,  Eq.~(\ref{eq:N15}) can also be used to predict, for example, the contact angle of a droplet of non-volatile liquids of any size.

\begin{figure}[htbp]
\begin{center}
\includegraphics[width=0.35\linewidth]{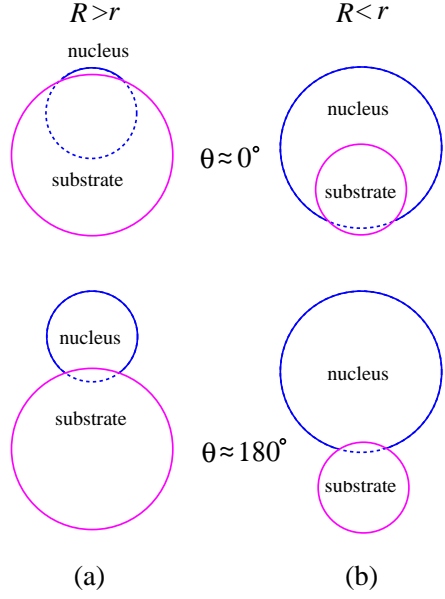}
\end{center}
\caption{
A schematic of the critical nucleus (blue sphere) on the spherical substrate (pink sphere) with its contact angle $\theta\approx 180^{\circ}$ and $\theta\approx 0^{\circ}$ when (a) $R>r$ and (b) $R>r$.  Although the critical nuclei for $r<R$ and $r>R$ look similar when $\theta\approx 180^{\circ}$, they are completely different for $\theta\approx 0^{\circ}$.   } 
\label{fig:N2}
\end{figure}

Figure~\ref{fig:N2} shows a typical shape of a nucleus on a hydrophilic surface with $\theta\approx 0^{\circ}$ and on a hydrophobic surface with $\theta\approx 180^{\circ}$ when the relative radius of the nucleus $r$ and that
of the substrate $R$ satisfies (a) $R>r$ and (b) $R<r$.  It is apparent that the three-phase contact line will stay on the upper hemisphere when $R>r$, and will cross the equator and move from the upper to lower hemisphere when the contact angle crosses $\phi=90^{\circ}$ when $R<r$.  Then, the sign of the last term in Eq.~(\ref{eq:N15}) will change as the line moves from the upper to the lower hemisphere.  Note that by increasing the contact angle, the three-phase contact-line length will increase in the lower hemisphere and will decrease in the upper hemisphere.

Equation (\ref{eq:N13}) has also been derived from the general theory of differential geometry using the geodesic curvature~\cite{Guzzardi2007}.  In the limit of an infinite substrate radius ($R\rightarrow\infty$), we can recover the modified Young equation~\cite{Navascues1981},
\begin{equation}
\Delta\sigma+\sigma_{\rm lv}\cos\theta+\frac{1}{r\sin\theta}\tau=0,
\label{eq:N16}
\end{equation}
or
\begin{equation}
\sigma_{\rm lv}\cos\theta=\sigma_{\rm lv}\cos\theta_{0}-\frac{\tau}{r\sin\theta},
\label{eq:N17}
\end{equation}
from Eq.~(\ref{eq:N13}) similar to Eq.~(\ref{eq:N15})  for a nucleus on a flat substrate.

\subsection{B. Nucleus on a wall of a concave spherical cavity}

\begin{figure}[htbp]
\begin{center}
\includegraphics[width=0.35\linewidth]{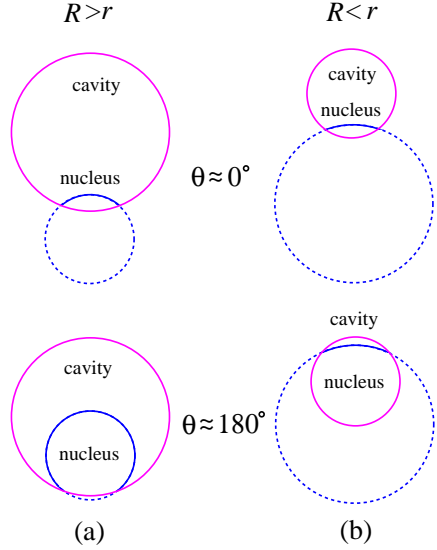}
\end{center}
\caption{
A schematic of the critical nucleus (blue sphere) in a spherical cavity (pink sphere) with its contact angle $\theta\approx 180^{\circ}$ and $\theta\approx 0^{\circ}$.  Although the critical nuclei for $r<R$ and $r>R$ look similar when $\theta\approx 0^{\circ}$, they are completely different for $\theta\approx 180^{\circ}$.   } 
\label{fig:N3}
\end{figure}

It is also possible to study the contact angle of a nucleus nucleated on a concave spherical substrate (cavity), such as that shown in Fig.~\ref{fig:N1}(b).  Figure~\ref{fig:N3} shows the typical configuration of a nucleus within a cavity.  We consider a semi-spherical nucleus with radius $r$ nucleated on a spherical cavity with radius $R$, as shown in Fig.~\ref{fig:N1}(b).  Using the geometrical angles $\phi$ and $\psi$, we obtain 
\begin{equation}
A_{\rm lv}=2\pi r^{2}\left(1-\cos\psi\right)=\pi r\frac{R^{2}-\left(C-r\right)^2}{C},
\label{eq:N18}
\end{equation}
and
\begin{equation}
A_{\rm sl}=2\pi R^{2}\left(1-\cos\phi\right)=\pi R\frac{r^{2}-\left(C-R\right)^2}{C}
\label{eq:N19}
\end{equation}
for the surface area of the liquid-vapor and liquid-solid interface, respectively.  The length of the three-phase contact line remains the same as that given by Eq.~(\ref{eq:N5}).  Now the distance $C$ between the two centers of the spheres with radii $R$ and $r$ are related through (Fig.~\ref{fig:N1}(b))
\begin{eqnarray}
C\sin\phi &=& r\sin\theta,\;\;\;C\sin\psi=R\sin\theta, \nonumber \\
C\cos\phi &=& R+r\cos\theta,\;\;\;\; C\cos\psi=r+R\cos\theta,
\label{eq:N20}
\end{eqnarray}
and the distance $C$ is given by
\begin{equation}
C^2=R^{2}+r^{2}+2Rr\cos\theta.
\label{eq:N21}
\end{equation}
The volume of the nucleus becomes
\begin{equation}
V = \frac{\pi}{12C}\left(C-R-r\right)^{2}\left[-3\left(R-r\right)^2+2C\left(R+r\right)+C^{2}\right],
\label{eq:N22}
\end{equation}
which again reduces to the more intuitive formula similar to Eq.~(\ref{eq:N9}), where the subtraction of two hemispherical volumes in Eq.~(\ref{eq:N9}) is replaced by the addition of two hemispherical volumes.

The contact angle is determined by minimizing the Helmholtz free energy in Eq.~(\ref{eq:N1}) under the condition of the constant nucleus volume given by Eq.~(\ref{eq:N22}).  Using the same procedure as that used in the previous subsection, we arrive at 
\begin{equation}
\Delta \sigma=+\frac{R^{2}+r^{2}-C^{2}}{2Rr}\sigma_{\rm lv}-\frac{C^{2}+R^{2}-r^{2}}{R\sqrt{2C^{2}\left(R^{2}+r^{2}\right)-\left(R^{2}-r^{2}\right)^{2}-C^{4}}}\tau,
\label{eq:N23}
\end{equation}
which gives the contact angle $\theta$ through
\begin{equation}
\Delta\sigma+\sigma_{\rm lv}\cos\theta+\frac{R+r\cos\theta}{Rr\sin\theta}\tau=0
\label{eq:N24}
\end{equation}
from Eq.~(\ref{eq:N21}) instead of Eq.~(\ref{eq:N7}).  Then, Eq.~(\ref{eq:N24}) can also be written as Eq.~(\ref{eq:N15}).  Therefore, the role of the line tension changes at $\phi=90^{\circ}$ again for a small cavity with $R<r$ (Fig.~\ref{fig:N3}(b)) even when the nucleus is confined within a spherical cavity.

Equation~(\ref{eq:N15}) suggests that the limit $\phi\rightarrow 0^{\circ}$ which corresponds to the limit $\theta\rightarrow 0^{\circ}$ is unphysical (see also Eqs.~(\ref{eq:N13}) and (\ref{eq:N24})).  Figures \ref{fig:N2} and \ref{fig:N3}, and Eqs.~(\ref{eq:N7}) and (\ref{eq:N21}) indicate that the nucleus on a convex substrate with $R>r$ and in a concave cavity disappear when $\theta\rightarrow 0^{\circ}$. In fact, the substrate must be covered by a microscopic thin liquid layer before the nucleus and the perimeter of the contact line disappear. This is similar to the limit $\theta_{0}\rightarrow 0^{\circ}$ of the classical Young equation in Eq.~(\ref{eq:N14}), which does not imply that the nucleus disappears. Rather, it implies that the bare substrate with the surface energy $\sigma_{\rm sv}$ will be covered by a thin wetting layer of the free energy $\sigma_{\rm sl}+\sigma_{\rm lv}$.  This wetting layer can be macroscopic only on a flat substrate since the liquid-vapor interface is flat and the Laplace pressure of curved surface vanishes.

\section{\label{sec:sec3} 3. Line tension effect on the critical nucleus of heterogeneous nucleation}

\subsection{A. Nucleus on a convex spherical substrate}

In the previous section, we studied the line-tension effect on the contact angle by minimizing the Helmholtz free energy.  Because we used the canonical ensemble of fixed volume, the nucleus is stable, and the results will be applicable to non-volatile liquids.  It is possible to study the critical nucleus of heterogeneous nucleation of a volatile liquid.  In this case, the metastable nucleus must be studied, and we must use the grand canonical ensemble.  Now, the nucleus is not simply a droplet but a metastable and transient critical nucleus of heterogeneous nucleation.

The critical nucleus is determined by maximizing the Gibbs free energy of formation~\cite{Navascues1981}
\begin{equation}
\Delta G=\Delta F-\Delta p V,
\label{eq:N25}
\end{equation}
where $\Delta p$ is the excess vapor pressure of the oversaturated vapor relative to the saturated pressure.  The Helmholtz free energy $\Delta F$ is given by Eq.~(\ref{eq:N1}), and the nucleus volume is given by Eq.~(\ref{eq:N8}).  By extremizing Eq.~(\ref{eq:N25}) using the formula
\begin{equation}
\frac{d\Delta G}{dr}=\frac{\partial \Delta G}{\partial r}+\frac{\partial \Delta G}{\partial C}\frac{dC}{dr}=0
\label{eq:N26}
\end{equation}
similar to Eq.~(\ref{eq:N11}), and using Eq.~(\ref{eq:N12}), which corresponds to the minimization of the Helmholtz free energy $\Delta F$, we obtain an algebraic equation
\begin{equation}
\left(C-R+r\right)^{2}\left(C+R+r\right)^{2}\left(C^{2}-R^{2}-4Cr+r^{2}\right)\left(\Delta p r-2\sigma_{lv}\right)=0,
\label{eq:N27}
\end{equation}
from which we obtain the well-known formula 
\begin{equation}
r_{*}=\frac{2\sigma_{\rm lv}}{\Delta p}
\label{eq:N28}
\end{equation}
for the critical radius of CNT.  The critical radius of the nucleus does not depend on the contact angle $\theta$ or the radius $R$ of the spherical substrate.  Eq.~(\ref{eq:N28}) simply implies that the curvature $1/r$ at the top of the nucleus is determined by the Young-Laplace equation because the effect of the spherical substrate is confined only at the surface of the substrate.  Therefore, neither the surface tensions $\sigma_{\rm sl}$ and, $\sigma_{\rm lv}$ nor the line tension $\tau$ affect the critical radius.  Likewise, the curvature $1/R$ also does not matter. When the interaction between the nucleus and substrate is long-ranged and is represented by the disjoining pressure, we will have a correction even on top of the nucleus~\cite{Iwamatsu2011}.

\begin{figure}[htbp]
\begin{center}
\includegraphics[width=0.80\linewidth]{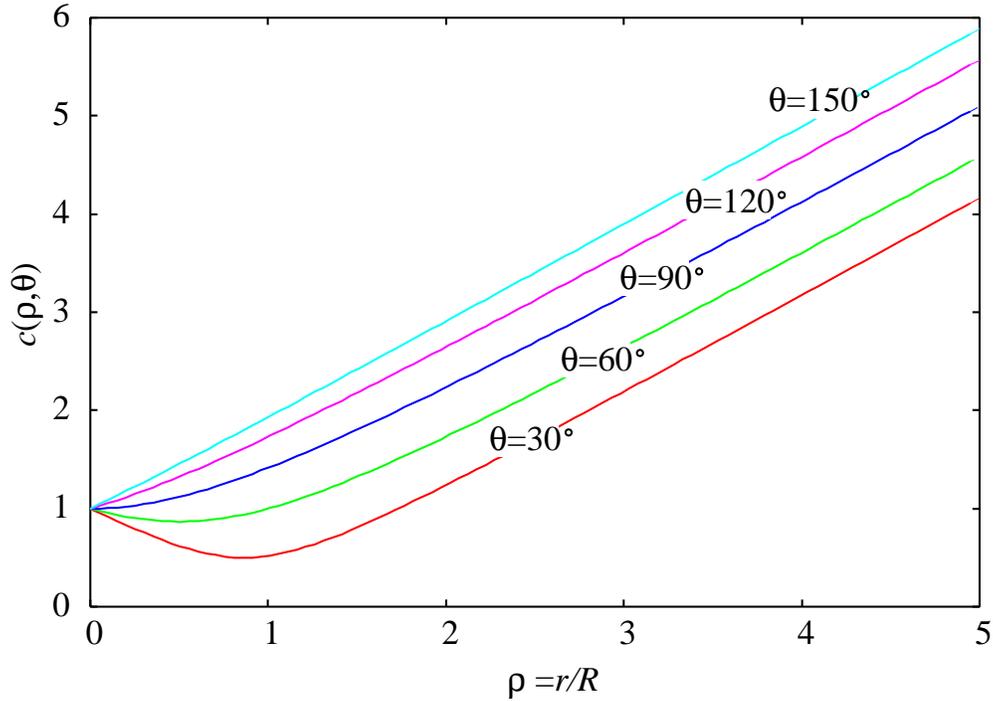}
\end{center}
\caption{
The distance $c$ between the center of the spherical substrate and spherical nucleus as a function of the parameter $\rho$ for various contact angles $\theta$.  }
\label{fig:N4}
\end{figure}

Once we know the radius of the critical nucleus $r_{*}$ from Eq.~(\ref{eq:N28}), we can determine the contact angle $\theta_{*}$ of the critical nucleus from Eq.~(\ref{eq:N13}) by inserting the critical radius $r_{*}$. However, we will continue to use the symbol $r$ and $\theta$ instead of $r_{*}$ and $\theta_{*}$ to represent the critical radius and corresponding contact angle to make mathematical expressions simpler.  Then, we can determine the distance $C$ between two spheres from Eq.~(\ref{eq:N7}) given by
\begin{equation}
c\left(\rho,\theta\right)=\sqrt{1+\rho^{2}-2\rho\cos\theta},
\label{eq:N29}
\end{equation}
where we have introduced the dimensionless parameters
\begin{equation}
c=C/R,\;\;\;\;\;\;\rho=r/R,
\label{eq:N30}
\end{equation}
which is the dimensionless length scaled by $R$.  Figure~\ref{fig:N4} plots the distance $c$ as a function of the relative size $\rho$ for various contact angles $\theta$.  Note that the limit $\rho\rightarrow 0$ corresponds to a flat substrate, and $\rho\rightarrow \infty$ corresponds to a point impurity and is equivalent to homogeneous nucleation.  Figure~\ref{fig:N4} shows that, when the contact angle approaches $\theta=180^{\circ}$ and the substrate is hydrophobic, the nucleation is homogeneous, and the distance $c$ is simply given by the sum of two spheres $1+\rho$.  However, when the contact angle is small ($\theta=30^{\circ}$), the distance $c$ reaches a minimum near $\rho\approx 1$.  

The shape of the nucleus on the spherical substrate is shown in Fig.~\ref{fig:N2}.  When the contact angle is almost $\theta\approx 0^{\circ}$, the nucleus with its radius $r$ smaller than the substrate radius $R$ ($r<R$) will disappear; however, that with a radius larger than the substrate ($r>R$) will wrap the spherical substrate (spherical seed) almost completely and will approach homogeneous nucleation.  However, when the contact angle is $\theta\approx 180^{\circ}$, the nucleus is almost the same as that of the homogeneous nucleation.

The critical radius of the nucleus in Eq.~(\ref{eq:N28}) is the same as that on the flat substrate even if the effect of line tension is included, because the effect of the substrate is confined only at the contact surface.  Inserting the critical radius $r$ given by Eq.~(\ref{eq:N28}) into Eq.~(\ref{eq:N25}) and using the generalized Young equation (Eq.~\ref{eq:N13}), we obtain the work of formation
\begin{equation}
\Delta G^{*}=\Delta G_{\rm vol}^{*}+\Delta G_{\rm lin}^{*},
\label{eq:N31}
\end{equation}
which consists of the volume term $\Delta G_{\rm vol}^{*}$ and the line term $\Delta G_{\rm lin}$ given by
\begin{eqnarray}
\Delta G_{\rm vol}^{*}&=&\frac{\pi\Delta p}{6}\left(R+2r-C\right)\left(C-R+r\right)^2, \nonumber \\
\Delta G_{\rm lin}^{*} &=&\frac{2\pi R\left(C-R+r\right)\left(R+r-C\right)\tau}{\sqrt{\left(C+R-r\right)\left(C-R+r\right)\left(R+r-C\right)\left(C+R+r\right)}}.
\label{eq:N32}
\end{eqnarray}
The former can be written as
\begin{equation}
\Delta G_{\rm vol}^{*} = f\left(\rho,\theta\right)\Delta G_{\rm homo}^{*}.
\label{eq:N33}
\end{equation}
Note that the limit $\rho\rightarrow 0$ corresponds to a flat substrate, and $\rho\rightarrow \infty$ represents a nucleus with a point impurity.  Therefore, the latter corresponds to homogeneous nucleation with the work of formation given by
\begin{equation}
\Delta G_{\rm homo}^{*}=\frac{2\pi}{3}r^{3}\Delta p.
\label{eq:N34}
\end{equation}
The generalized shape factor $f\left(\rho,\theta\right)$ is given by
\begin{equation}
f\left(\rho,\theta\right)=\frac{1}{4\rho^{3}}
\left(1+2\rho-\sqrt{1+\rho^{2}-2\rho\cos\theta}\right)
\left(-1+\rho+\sqrt{1+\rho^{2}-2\rho\cos\theta}\right)^{2},
\label{eq:N35}
\end{equation}
which reduces to the well-known shape factor originally derived by Fletcher~\cite{Fletcher1958} after tedious manipulation of algebra.  It also reduces to the shape factor~\cite{Turnbull1950,Navascues1981}
\begin{equation}
f\left(\rho\rightarrow 0,\theta\right)=\frac{\left(2+\cos\theta\right)\left(1-\cos\theta\right)^2}{4}
\label{eq:N36}
\end{equation}
for a flat substrate ($\rho\rightarrow 0$) and 
\begin{equation}
f\left(\rho\rightarrow \infty, \theta\right)=1,
\label{eq:N37}
\end{equation}
which corresponds to homogeneous nucleation.  Similarly, we can show
\begin{equation}
f\left(\rho,\theta\rightarrow 180^{\circ}\right)=1,
\label{eq:N38}
\end{equation}
which also corresponds to homogeneous nucleation as the substrate is super-hydrophobic with the contact angle $\theta=180^{\circ}$.  

\begin{figure}[htbp]
\begin{center}
\includegraphics[width=0.80\linewidth]{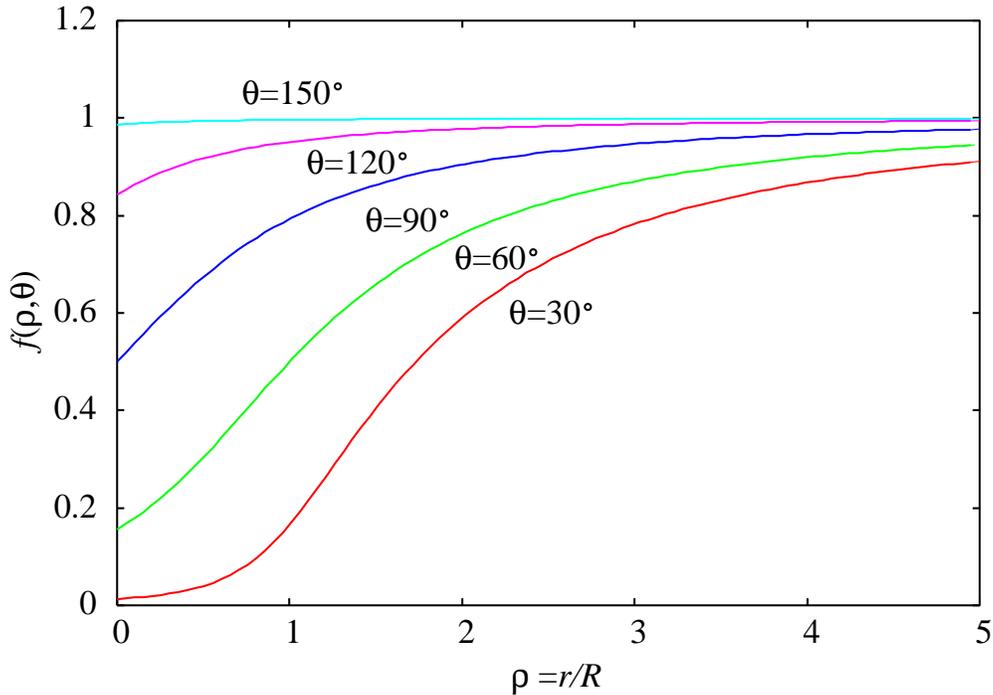}
\end{center}
\caption{
The generalized shape factor $f\left(\rho,\theta\right)$ on a spherical substrate as a function of the ratio $\rho$ for various $\theta$.  The limit $f\left(\rho\rightarrow 0,\theta\right)$ is given by Eq.~(\ref{eq:N36}).   } 
\label{fig:N5}
\end{figure}

Figure~\ref{fig:N5} shows the generalized shape factor $f\left(\rho,\theta\right)$ on a spherical substrate as the function of the ratio $\rho$ for several contact angles $\theta$.  When the radius $R$ is finite ($\rho>0$), the volume contribution $\Delta G_{\rm vol}$ is always larger than that on a flat substrate.  Therefore, the nucleation is more favorable on a flat substrate than on the spherical impurity.  In addition, because $f\left(\rho\rightarrow \infty,\theta\right)\rightarrow 1$, a spherical substrate whose radius $R$ is much smaller than the critical radius $r_{*}$ ($\rho\gg 1$) is not an active nucleation site as the work of formation does not differ from that of the homogeneous nucleation. The spherical substrate is a more effective nucleation site when the contact angle is smaller and the substrate is more hydrophilic. 

The line contribution $\Delta G_{\rm lin}^{*}$ in Eqs.~(\ref{eq:N31}) and (\ref{eq:N32}) can be written as
\begin{equation}
\Delta G_{\rm lin}^{*}=2\pi r\tau g\left(\rho,\theta\right)
\label{eq:N39}
\end{equation}
using the generalized shape factor for the line contribution
\begin{equation}
g\left(\rho,\theta\right)
=\frac{-1+\rho\cos\theta+\sqrt{1+\rho^{2}-2\rho\cos\theta}}
{\rho^{2}\sin\theta},
\label{eq:N40}
\end{equation}
which reduces to
\begin{equation}
g\left(\rho\rightarrow 0,\theta\right)=\frac{\sin\theta}{2}
\label{eq:N41}
\end{equation}
for a flat substrate~\cite{Navascues1981}, and
\begin{equation}
g\left(\rho\rightarrow \infty,\theta\right)=0,
\label{eq:N42}
\end{equation}
which corresponds to the homogeneous nucleation for which the three-phase contact line disappears.  Similarly we have
\begin{equation}
g\left(\rho,\theta\rightarrow 180^{\circ}\right)=0
\label{eq:N43}
\end{equation}
for the super-hydrophobic substrates ($\theta=180^{\circ}$) as the nucleation is homogeneous again as shown in Fig.~\ref{fig:N2}.   Equation (\ref{eq:N40}) can also be written simply as
\begin{equation}
g\left(\rho, \phi\right)=\frac{1-\cos\phi}{\rho\sin\phi}
\label{eq:N44}
\end{equation}
as a function of $\phi$ instead of $\theta$.  In contrast to Eq.~(\ref{eq:N15}), the sign will not change at $\phi=90^{\circ}$.

\begin{figure}[htbp]
\begin{center}
\includegraphics[width=0.80\linewidth]{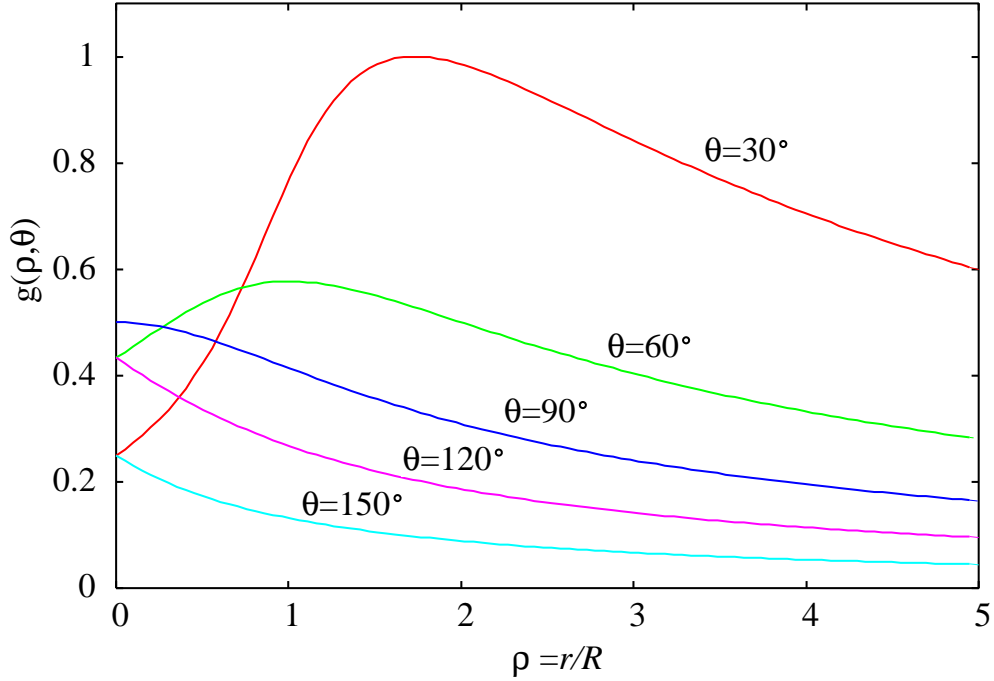}
\end{center}
\caption{
The generalized shape factor $g\left(\rho,\theta\right)$ on a spherical substrate for the line contribution as a function of the ratio $\rho$ for various $\theta$.  The limit $g\left(\rho\rightarrow 0,\theta\right)$ is given by Eq.~(\ref{eq:N41}).   } 
\label{fig:N6}
\end{figure}

Figure~\ref{fig:N6} shows the shape factor $g\left(\rho,\theta\right)$ for the line contribution as a function of the ratio $\rho$ for several contact angles $\theta$.  When the radius $R$ is finite ($\rho>0$), the line contribution $\Delta G_{\rm lin}$ or the shape factor  $g\left(\rho,\theta\right)$ is always smaller than that on a flat substrate if the substrate is hydrophobic ($\theta>90^{\circ}$).  Therefore, the line tension affects the nucleation barrier more effectively on a flat substrate than on a spherical substrate for a hydrophobic surface ($\theta>90^{\circ}$).  In contrast, the shape factor $g\left(\rho,\theta\right)$ becomes maximum when $\rho\approx 1-2$ when the substrate is hydrophilic ($\theta<90^{\circ}$).  Then, the line contribution $\Delta G_{\rm lin}$ becomes larger than that on a flat substrate.  The line tension is more effective for a spherical substrate with its radius comparable to the nucleus than for a flat substrate when the substrate is hydrophilic.

The nucleation barrier becomes higher ($\tau>0$) or lower ($\tau<0$) according to Eqs.~(\ref{eq:N31}) and (\ref{eq:N39}) when including the line contribution depending on the sign of the line tension compared with not including the line tension.  The line-tension effect is most effective on a flat substrate if the substrate is hydrophobic $\theta>90^{\circ}$.  The line-tension effect is most effective on a spherical substrate with its radius comparable to that of the critical nucleus if the substrate is hydrophilic $\theta<90^{\circ}$.

\subsection{B. Nucleus on a wall of a concave spherical cavity}

Next, we will consider heterogeneous nucleation on the inner wall of a spherical cavity.  Figure~\ref{fig:N3} shows a nucleus (droplet) in a spherical cavity when the contact angle is $\theta\approx 0^{\circ}$ and $\theta\approx 180^{\circ}$.  Similar to Fig.~\ref{fig:N2}, there are two scenarios with $r<R$ and $r>R$.  Although the nuclei for both cases are almost similar when $\theta\approx 0^{\circ}$, they are completely different when $\theta\approx 180^{\circ}$ in contrast to Fig.~\ref{fig:N2}.  Here, the nucleus is almost the same as that of the homogeneous nucleation when $r<R$, as observed in Fig.~\ref{fig:N2}.  However, the nucleus is different from that of the homogeneous nucleation when $r>R$ as the nucleus is only a small part of the nucleus because the space is confined within the small cavity.

By minimizing the free energy (\ref{eq:N25}), we obtain an algebraic equation
\begin{equation}
\left(C-R-r\right)^{2}\left(C+R-r\right)^{2}\left(C^{2}-R^{2}+4Cr+r^{2}\right)\left(\Delta p r-2\sigma_{lv}\right)=0,
\label{eq:N45}
\end{equation}
from which we obtain the Young-Laplace formula (\ref{eq:N28}) for the critical radius again.  

Once we know the radius of the critical nucleus $r_{*}$,  the contact angle $\theta_{*}$ of the critical nucleus will be determined from Eq.~(\ref{eq:N24}) by inserting the critical radius $r_{*}$ determined from Eq.~(\ref{eq:N28}).  Then, we can determine the distance $C$ between two spheres using Eq.~(\ref{eq:N21}) as follows:
\begin{equation}
c=\sqrt{1+\rho^{2}+2\rho\cos\theta},
\label{eq:N46}
\end{equation}
where we have introduced dimensionless parameters of Eq.~(\ref{eq:N30}) and continued to use $r$ and $\theta$ instead of $r_{*}$ and $\theta_{*}$ for the critical nucleus. The distance $c$ as a function of the parameter $\rho$ is given in Fig~\ref{fig:N4} by replacing $\theta\rightarrow 180^{\circ}-\theta$.

Therefore, on a concave spherical substrate, the critical radius of nucleus is also the same as that on a flat substrate even if the effect of the line tension is included because the effect of the substrate, including that of the line tension, is confined only to the contact surface.  Inserting the critical radius $r_{*}$ into Eq.~(\ref{eq:N25}) and using the generalized Young equation (Eq.~(\ref{eq:N24})), we can obtain the work of formation (Eq.~(\ref{eq:N31})) of the nucleus within a spherical cavity.

Inserting the critical radius $r$ into Eq.~(\ref{eq:N25}) and using the generalized Young equation (\ref{eq:N23}) we obtain the work of formation  given by
\begin{eqnarray}
\Delta G_{\rm vol}^{*}&=&\frac{\pi\Delta p}{6}\left(C-R+2r\right)\left(C-R-r\right)^2, \nonumber \\
\Delta G_{\rm lin}^{*} &=&\frac{2\pi R\left(C-R+r\right)\left(R+r-C\right)\tau}{\sqrt{\left(C+R-r\right)\left(C-R+r\right)\left(R+r-C\right)\left(C+R+r\right)}},
\label{eq:N47}
\end{eqnarray}
which can be written in analogous way as Eqs.~(\ref{eq:N33}) and (\ref{eq:N39}).

The volume term $\Delta G_{\rm vol}^{*}$  can also be written as Eq.~(\ref{eq:N33}), and 
the limit $\rho\rightarrow 0$ corresponds to a flat substrate and $\rho\rightarrow \infty$ represents a nucleus confined within a small cavity.  The nucleus with $\theta\approx 180^{\circ}$ and $\rho<1$ (Fig.~\ref{fig:N3}, bottom left) corresponds to the homogeneous nucleation with the work of formation given by Eq.~(\ref{eq:N34}).  That with $\theta\approx 180^{\circ}$ and $\rho\rightarrow \infty$ (Fig.~\ref{fig:N3}, bottom right) also corresponds to a spherical nucleus, however, its volume is confined within an infinitesimally small spherical cavity.

The generalized shape factor $f\left(\rho,\theta\right)$ for a nucleus in a cavity is now given by
\begin{equation}
f\left(\rho,\theta\right)=\frac{1}{4\rho^{3}}
\left(-1+2\rho+\sqrt{1+\rho^{2}+2\rho\cos\theta}\right)
\left(1+\rho-\sqrt{1+\rho^{2}+2\rho\cos\theta}\right)^{2},
\label{eq:N48}
\end{equation}
which again reduces to the well-known shape factor~\cite{Fletcher1958,Navascues1981,Qian2009} in Eq.~(\ref{eq:N36}) for a flat substrate ($\rho\rightarrow 0$), and 
\begin{equation}
f\left(\rho\rightarrow \infty,\theta\right)=0
\label{eq:N49}
\end{equation}
because the nucleus is confined within an infinitesimally small spherical cavity. We can also show that
\begin{equation}
f\left(\rho<1,\theta\rightarrow 180^{\circ}\right)=1,
\label{eq:N50}
\end{equation}
which corresponds to homogeneous nucleation as the cavity is large enough to accommodate the entire nucleus and the substrate is super-hydrophobic with the contact angle $\theta=180^{\circ}$.  

\begin{figure}[htbp]
\begin{center}
\includegraphics[width=0.80\linewidth]{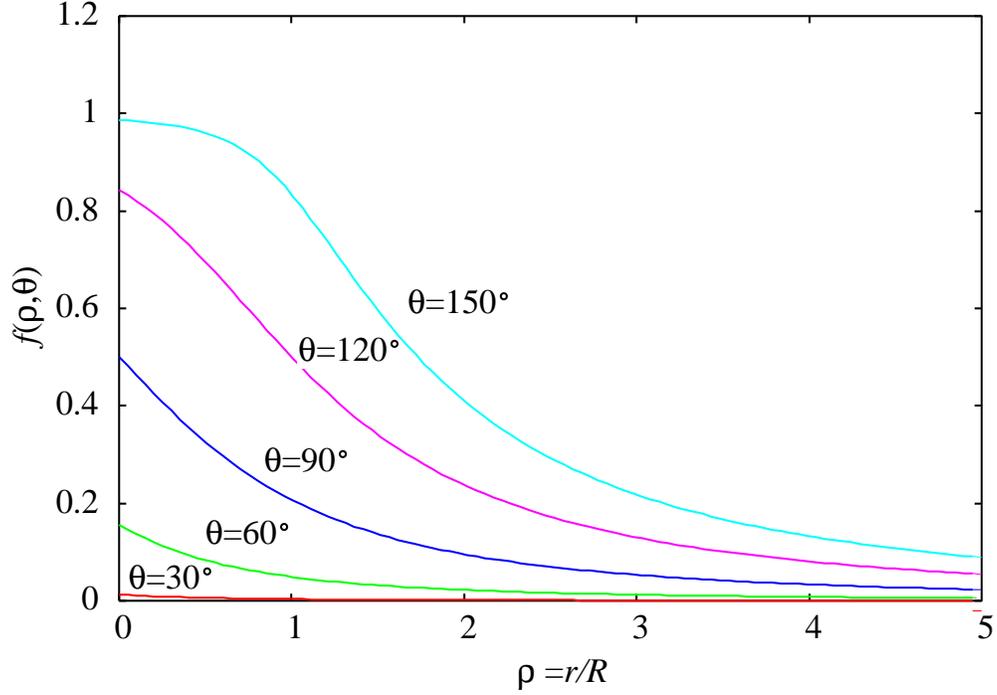}
\end{center}
\caption{
The generalized shape factor $f\left(\rho,\theta\right)$ in a spherical cavity as a function of the ratio $\rho$ for various $\theta$.  The limit $f\left(\rho\rightarrow 0,\theta\right)$ is given by Eq.~(\ref{eq:N36}).   } 
\label{fig:N7}
\end{figure}

Figure~\ref{fig:N7} shows the generalized shape factor $f\left(\rho,\theta\right)$ in a spherical cavity as a function of the ratio $\rho$ for several contact angles $\theta$.  When the radius $R$ is finite ($\rho>0$), the volume contribution $\Delta G_{\rm vol}$ is always smaller than that on a flat substrate.  Therefore, the nucleation is more favorable in a spherical cavity than on a flat substrate as the nucleation barrier will be lower.  In addition,  because $f\left(\rho\rightarrow \infty,\theta\right)\rightarrow 0$, a spherical cavity whose radius $R$ is much smaller than the critical radius $r_{*}$ ($\rho\gg 1$) is an active nucleation site as the work of formation approaches zero to achieve athermal nucleation.

The line contribution $\Delta G_{\rm lin}^{*}$ in Eq.~(\ref{eq:N31}) can also be written as Eq.~(\ref{eq:N39}).  The generalized shape factor of the line contribution for a nucleus within a spherical cavity is given by
\begin{equation}
g\left(\rho,\theta\right)
=\frac{-1-\rho\cos\theta+\sqrt{1+\rho^{2}+2\rho\cos\theta}}
{\rho^{2}\sin\theta},
\label{eq:N51}
\end{equation}
which reduces to
\begin{equation}
g\left(\rho\rightarrow 0,\theta\right)=\frac{\sin\theta}{2}
\label{eq:N52}
\end{equation}
for the flat substrate~\cite{Navascues1981} again, and
\begin{equation}
g\left(\rho\rightarrow \infty,\theta\right)=0,
\label{eq:N53}
\end{equation}
which corresponds to an infinitesimally small nucleation, for which the three-phase contact line vanishes.  Similarly, we have
\begin{equation}
g\left(\rho,\theta\rightarrow 180^{\circ}\right)=0
\label{eq:N54}
\end{equation}
for super-hydrophobic substrates ($\theta=180^{\circ}$) as the nucleation is homogeneous again ($\rho<1$) or the nucleus completely fills the cavity ($\rho>1$), as demonstrated in Fig.~\ref{fig:N3}.   Equation (\ref{eq:N51}) can also be written simply as Eq.~(\ref{eq:N44}) as a function of $\phi$ instead of $\theta$.  

\begin{figure}[htbp]
\begin{center}
\includegraphics[width=0.80\linewidth]{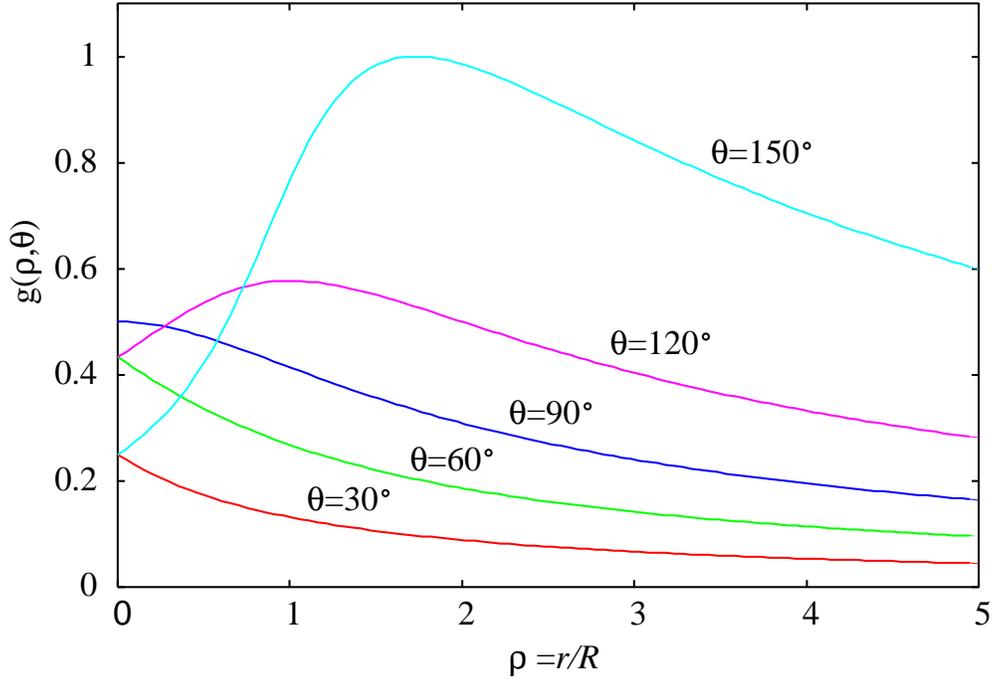}
\end{center}
\caption{
The generalized shape factor $g\left(\rho,\theta\right)$ in a spherical cavity as a function of the ratio $\rho$ for various $\theta$.  The limit $g\left(\rho\rightarrow 0,\theta\right)$ is given by Eq.~(\ref{eq:N52}).   } 
\label{fig:N8}
\end{figure}

\begin{figure}[htbp]
\begin{center}
\includegraphics[width=0.30\linewidth]{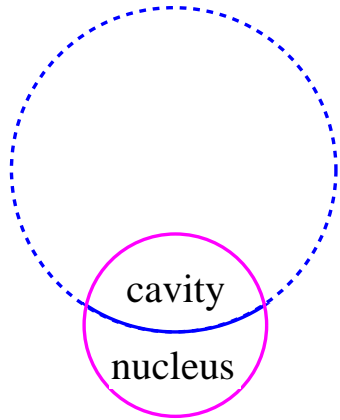}
\end{center}
\caption{
A nucleus with a concave liquid-vapor interface within a cavity when the vapor is undersaturated.     } 
\label{fig:N9}
\end{figure}

It can be observed that the generalized shape factor, Eq.~(\ref{eq:N51}), in a spherical cavity can be obtained by replacing $\theta\rightarrow 180^{\circ}-\theta$ in Eq.~(\ref{eq:N40}) and in Fig.~\ref{fig:N6} of a spherical substrate.  Therefore, the hydrophobicity and hydrophilicity simply change their roles from sphere to cavity.  However, for the sake of completeness, in Fig~\ref{fig:N8}, we show the shape factor $g\left(\rho,\theta\right)$ in Eq.~(\ref{eq:N51}) for the nucleus in a spherical cavity.  Now, the effects of hydrophobicity and hydrophilicity are exchanged compared to those shown in Fig.~\ref{fig:N6}.  Therefore, the line tension affects the nucleation barrier most effectively on a flat substrate than on a spherical substrate for a hydrophilic surface ($\theta<90^{\circ}$) in a cavity.  In contrast, the shape factor $g\left(\rho,\theta\right)$ becomes larger than that on a flat substrate on a hydrophobic surface ($\theta>90^{\circ}$).  The line tension is more effective for a spherical cavity with its radius being comparable to the nucleus than for a flat substrate when the substrate is hydrophobic.  Furthermore, the effect of line tension can be more important in spherical cavity than on spherical substrate because the volume term represented by the shape factor  $f\left(\rho,\theta\right)$ in a spherical cavity will always be smaller that on a flat substrate as shown in Fig.~\ref{fig:N7}.

Although we have considered a nucleus with a convex spherical liquid-vapor surface, it is possible to consider a nucleus or wetting layer with a concave surface, as shown in Fig.~\ref{fig:N9}.  In this case, however, we must consider an undersaturated vapor with $\Delta p<0$ and study a bubble nucleus rather than a liquid nucleus.  Then, the description for the liquid nucleus in the cavity is applicable.

\section{\label{sec:sec5}4. Conclusion}
In this paper, we have studied an old problem of the line-tension effect on a lens-shaped nucleus (liquid droplet)  on a spherical substrate and in a spherical cavity.  By minimizing the Helmholtz free energy, we reproduced the generalized Young formula determined by previous authors~\cite{Guzzardi2007,Hienola2007} to determine the contact angle.  Then, we studied the work of formation (nucleation barrier) of the critical nucleus by maximizing the Gibbs free energy together with the generalized Young formula.  We determined that the work of formation consists of the usual volume contribution and the line contribution due to the line tension.  The former for the nucleus on a spherical substrate coincides with that obtained originally by Fletcher~\cite{Fletcher1958}.   The volume contribution suggests that the convex spherical surface will not enhance the heterogeneous nucleation as the nucleation approaches homogeneous nucleation.  However, the concave spherical surface will enhance the heterogeneous nucleation, as the free energy barrier will be lower.  The line contribution depends sensitively on the hydrophobicity and hydrophilicity of the substrate.  On a hydrophilic spherical substrate, the line contribution can be maximized when the radius of the substrate is finite.  However, it can be maximized when the radius is finite for a hydrophobic cavity.  

Although, we have used a spherical lens model and assumed that the liquid-solid interaction is represented by the short-ranged contact interaction represented by the surface tension,  it is possible to include long-ranged liquid-solid interaction using the concept of disjoining pressure.  Several studies~\cite{Hage1984,Kuni1996,Tsekov2000,Iwamatsu2013} have already examined the effect of the long-range force.  However, most studies have focused on the problem of a spherical nucleus around a spherical substrate without the three-phase contact line~\cite{Kuni1996,Bieker1998,Tsekov2000,Bykov2006} rather than a lens-shaped nucleus with the three-phase contact line.  Although there are numerous studies on lens-shaped nuclei on flat substrates when the long-ranged disjoining pressure exists~\cite{Yeh1999,Zhang2002a,Starov2004,Iwamatsu2011}, investigations of the lens-shaped nuclei on spherical substrates are scarce and remain qualitative as the result is only numerical~\cite{Dobbs1992}.  In additon, we have assumed that the liquid-vapor interface is sharp.  There are also several studies~\cite{Padilla2001,Ghosh2013} that include diffuse interfaces.  Our results in this paper can serve as a guide in the development of more realistic models of lens-shaped nuclei on curved surfaces to include the disjoining pressure and diffuse interface effect.


\begin{suppinfo}
Mathematical detail of the derivation of some formulas. 
\end{suppinfo}



\begin{tocentry}

\begin{center}
\includegraphics[width=0.61\linewidth]{figN1.eps}
\end{center}
\end{tocentry}

\end{document}